\documentclass[fleqn,usenatbib]{mnras}

\usepackage{newtxtext,newtxmath}

\usepackage[T1]{fontenc}
\usepackage{ae,aecompl}

\usepackage{graphicx}	
\usepackage{amsmath}	
\usepackage{amsmath, epsfig,natbib}
\usepackage{multicol}
\usepackage{color,ulem}
\usepackage{newtxtext,newtxmath}
\definecolor{webgreen}{rgb}{0,.5,0}
\definecolor{webbrown}{rgb}{.6,0,0}

\newcommand{\kms}{\mbox{$\>{\rm km\, s^{-1}}$}}
\newcommand{\kmskpcGyr}{\mbox{$\>{\rm km\, s^{-1}\, kpc^{-1}\, Gyr^{-1}}$}}

\newcommand{\kpc}{\mbox{$\>{\rm kpc}$}} 
\newcommand{\pixel}{\mbox{$\>{\rm pixel}$}} 

\newcommand{\Gyr}{\mbox{$\>{\rm Gyr}$}}
\newcommand{\Myr}{\mbox{$\>{\rm Myr}$}}

\newcommand{\Msun}{\>{\rm M_{\odot}}}

\newcommand\degrees{^\circ}

\title[multiple pattern speeds in long peanut bar]
{Multiple pattern speeds in a long peanut-shaped bar in a simulated galaxy}

\author[Vynatheya et al.]{
Pavan Vynatheya,$^{1,2}$\thanks{E-mail: pavanvyn@mpa-garching.mpg.de}
Kanak Saha$^{3}$ \thanks{E-mail: kanak@iucaa.in},
Soumavo Ghosh$^{4,3}$ \thanks{E-mail: soumavo.ghosh@iiap.res.in}\\
$^{1}$Max-Planck-Institut für Astrophysik, Karl-Schwarzchild-Str. 1, 85748 Garching bei München, Germany\\
$^{2}$Indian Institute of Science Education and Research Kolkata, Mohanpur, Nadia - 741246, West Bengal, India\\
$^{3}$Inter-University Center for Astronomy and Astrophysics, Post Bag 4, Ganeshkhind, Pune - 411007, India\\
$^{4}$ Indian Institute of Astrophysics, Koramangala II Block, Bengaluru 560034, India\\
}

\date{Accepted XXX. Received YYY; in original form ZZZ}

\pubyear{2021}

\begin{document}
\label{firstpage}
\pagerange{\pageref{firstpage}--\pageref{lastpage}}
\maketitle

\begin{abstract}
A significant fraction of barred spiral galaxies exhibits peanut/X-shaped structures in their central regions. Bars are known to rotate with a single pattern speed, and they eventually slow down over time due to the dynamical friction with the surrounding dark matter halo. However, the nature of the decay in pattern speed values and whether all peanut bars rotate with a single pattern speed remain to be investigated. 

Using $N$-body simulation of a collisionless stellar disc, we study the case of a long bar with a three-dimensional peanut structure prominent in both edge-on and face-on projections. We show that such a bar possesses three distinct peaks in the $m=2$ Fourier component. Using the Tremaine-Weinberg method, we measure the pattern speeds and demonstrate that the three regions associated with the three peaks rotate with different pattern speeds. The inner region, which is the core of the peanut, rotates slower than the outer regions. In addition, the pattern speed of the inner bar also decays faster than the outer bar with a decay timescale of 4.5 Gyr for the inner part and $\sim 12.5$~Gyr for the outer parts. This is manifested as a systematic offset in density and velocity dispersion maps between the inner and outer regions of the long peanut bar. We discuss the importance of our findings in the context of bar dynamics.  
\end{abstract}



\section{Introduction}
\label{sec:intro}

Bars are one of the most common $m=2$ non-axisymmetric patterns in disc galaxies in the local Universe \citep[e.g., see][]{Eskridgeetal2000,Whyteetal2002,Aguerrietal2009,Mastersetal2011}, where $m$ is the azimuthal Fourier component of the disc light distribution. In observations, bars characterised by a peak in their ellipticity profiles associated with a nearly constant position angle in the radial range containing the peak \citep[e.g., ][]{ErwinSparke2003,Laurikainenetal2007,Kruketal2018}. Bars are more prominent in late-type disc galaxies \citep{NairandAbraham2010,Kruketal2018}. As one scans the morphological space (or the Hubble Tuning Fork diagram) from the late-type to the early-type disc galaxies, strong bars become scarce. The incidence of strong bars in the Lenticulars (S0) is at its minimum \citep[e.g., see][]{Laurikainenetal2009,Butaetal2010,NairandAbraham2010,Barway2011,Gaoetal2018}.  

Much of our current understanding of the dynamics of barred galaxies have been gleaned from the numerical simulations, where an axisymmetric disc galaxy forms a bar spontaneously \citep[e.g., see][]{Milleretal1970,Hohl1971,CombesandSanders1981,SellwoodandWilkinson1993,Athanassoula2003,Berentzenetal2006,Villa-Vargasetal2010,Sahaetal2010,Athanassoula2013,Sellwood2013,Sahaetal2018}. As shown in numerous simulations, a bar is known to rotate with a single pattern speed which marks a well-defined corotation radius ($R_{\rm CR}$) in the galaxy. The corotation radius divides the whole galaxy into two dynamically distinct parts: inside the corotation, where the stars rotate faster than the bar and outside the corotation, where the bar rotates faster than the background disc stars. In fact, in several barred galaxies, the ring present just outside the bar is understood to be associated with the corotation resonance \citep{ButaandCombes1996,Laurikainenetal2011,Buta2017}. Careful, accurate measurement of bar pattern speed is therefore extremely crucial to grasp the dynamics of barred galaxies. 

With the advent of integral field spectroscopy (hereafter IFS) and thanks to several publicly available IFS data such as CALIFA \citep{califa2011}, SAMI \citep{Croometal2012}, MaNGA \citep{Bundyetal2015}, it is now possible to measure the bar pattern speed in many barred galaxies by employing the Tremaine-Weinberg method \citep{TremaineWeinberg1984}. One of the underlying \textit{assumptions} while applying the Tremaine-Weinberg method to the stellar kinematic data from IFS surveys is that the bar rotates with a \textit{single} pattern speed \citep[e.g., see in][]{Aguerrietal2015,Cuomoetal2019,Guoetal2019,Williamsetal2021}. However, it remains elusive, to date, whether a bar, especially when it is large-scale and has developed a strong boxy/peanut structure, rotates coherently with a single pattern speed. 

Numerical simulations of isolated galaxies, where bars forms spontaneously, are an ideal and indispensable tool to address such a question \citep[e.g.,][]{CombesandSanders1981,SellwoodandWilkinson1993,Athanassoula2003}. After a bar forms in the simulation, it goes through morphological and dynamical changes. In simulations with a live dark matter halo, a bar grows in size and mass via transferring a fraction of the disc angular momentum to the dark matter halo and thereby trapping more and more particles in the bar \citep[e.g.,][]{TremaineandWeinberg1984,HernquistWeinberg1992,DebattistaSellwood2000, Athanassoula2002,SellwoodandDebattista2006,Dubinskietal2009,Sahaetal2012,SahaNaab2013}. As the bar becomes massive, it undergoes a buckling instability phase and forms a peanut bulge \citep{Combesetal1990,Rahaetal1991,Martinez-Valpuestaetal2006,Athanassoula2016,Sahaetal2018}. During the bar's evolution, its pattern speed is known to decrease due to the dynamical friction with the surrounding dark matter halo \citep{Weinberg1985,DebattistaSellwood1998,DebattistaSellwood2000,SahaNaab2013,Athanassoula2014}. However, there is no clear consensus to date on the rate at which a bar slows down and whether the slow-down follows an exponential or a power-law profile.  

In this paper, we investigate the particular case of a long bar with a prominent face-on peanut \citep{Sahaetal2018} developed in an isolated galaxy model using an N-body simulation. The inner core of this bar has a strong peanut, and the outer part has a distinct ansae-like structure \citep{Martinez-Valpuestaetal2007}. We follow the evolution of this bar in isolation and investigate the temporal evolution of the bar pattern speed using the Tremaine-Weinberg method \citep{TremaineWeinberg1984}. One advantage of the Tremaine-Weinberg method is that this allows an explicit measurement of the pattern speed from a single snapshot. We discuss the uncertainties associated with our pattern speed measurement and the nature of the bar's rotation in the galaxy model. 

\noindent The rest of the paper is organised as follows:
Section~\ref{sec:simu_setup} provides a brief description of the simulation set up. Section~\ref{sec:tw_method} is devoted to the Tremaine-Weinberg method and its application to our galaxy model. Section~\ref{sec:multiple_structure} describes the evolution of the bar and the formation of the complex structure within. Section~\ref{sec:pattern_speed_measurement} provides the pattern speed measurement and its temporal evolution. Section~\ref{sec:impact_patternspeed} discusses the fingerprints of different pattern speeds within the bar region. Sections~\ref{sec:discussion} and~\ref{sec:conclusion} contain the discussion and the main findings of this work, respectively.

\section{Details of the simulation set-up}
\label{sec:simu_setup}
%
For this work, we consider an $N$-body simulation of a collisionless stellar disc that harbours a long bar with a  prominent \textit{three-dimensional boxy/peanut} (b/p)-shaped structure. The details of the initial set-up and the evolution of the b/p structure has been described in \citet[see `sim2' model]{Sahaetal2018}. For the sake of completeness, we briefly mention the initial set-up of the simulation.

The initial equilibrium model consists of a disc, a (small) classical bulge, and a dark matter halo. Each component is modelled using a distribution function and are all \textit{live}; thereby allowing them to interact with each other \citep[for details, see][]{KuijkenDubinski1995,Sahaetal2012}. The initial radial surface density of the stellar disc follows an exponential fall-off and has a $sech^2$ profile along the vertical direction. The classical bulge is modelled with a cored King profile \citep{King1966} whereas the dark matter distribution is modelled as a cored halo \citep{Evans1993}. The masses of the stellar disc, the dark matter halo, and the bulge are $1.67 \times 10^{10} \Msun$, $6.5 \times 10^{10} \Msun$, and $0.022 \times 10^{10} \Msun$, respectively \citep[see Table.~1 in][]{Sahaetal2018}. A total of $3.7 \times 10^6$ particles have been used, with $0.5\times 10^6$ in the bulge component, $1.2 \times 10^6$ in the disc, and $2.0\times 10^6$ in the dark matter component. The simulation was run using {\sc{Gadget-1}} \citep{Springeletal2001} with a tolerance parameter $\theta_{\rm tol} =0.7$ and an integration time-step $\sim 1.8 \Myr$ \citep[for details see][]{Sahaetal2018}. The initial Toomre $Q$ for this model is less than 1 in nearly all radii \citep[see Fig.~1 in][]{Sahaetal2018}. The simulation was evolved for $\sim 13.2 \Gyr$, with a time difference of $60 \Myr$ between two consecutive snapshots. We use the pixel as the unit of length throughout this paper with $1 \pixel = 0.23 \kpc$, unless mentioned explicitly. We refer to this model as \textit{3-D peanut model} throughout this paper.

\section{Bar pattern speed measurement -- Tremaine-Weinberg (TW) method}
\label{sec:tw_method}
Here, we briefly describe the Tremaine-Weinberg method (hereafter TW method) used to measure the pattern speed of the bar present in our model. The TW method, initially devised by \citet{TremaineWeinberg1984}, is a method/technique for deriving the bar pattern speed by utilising \textit{only} the line-of-sight surface brightness distribution and the velocity field of a tracer, and hence is model-independent. The underlying assumptions associated with this method are (a) the galactic disc is flat (b) the tracer obeys the continuity equation, and (c) the bar rotates like a rigid body with a definite pattern speed ($\Omega_{\rm p}$). Under these assumptions and after some algebraic simplifications, the continuity equation becomes
\begin{equation}
\Omega_{\rm p} \sin i \int_{- \infty}^{\infty} \Sigma(X, Y) X dX = \int_{- \infty}^{\infty} \Sigma(X, Y) V_{los} (X, Y) dX\,,
\end{equation}
\noindent where $i$ is the angle of inclination, and $V_{los}$ is the line-of-sight velocity. $(X, Y)$ are the coordinates in the galactic plane so that $(X,Y) = (x,y \cos{i})$ where $(x, y)$ are the coordinates in the sky plane, as viewed by an observer \citep[for details see][]{TremaineWeinberg1984}. If the angle of inclination ($i$) is known, then the pattern speed ($\Omega_{\rm p}$) of the bar is determined as 
\begin{equation}
\Omega_{\rm p} \sin i= \frac{\int_{- \infty}^{\infty} \Sigma(X, Y) V_{los} (X, Y) dX}{\int_{- \infty}^{\infty} \Sigma(X, Y) X dX}\,.
\label{eq:tw_ori}
\end{equation}

\begin{figure*}
\includegraphics[width=\linewidth]{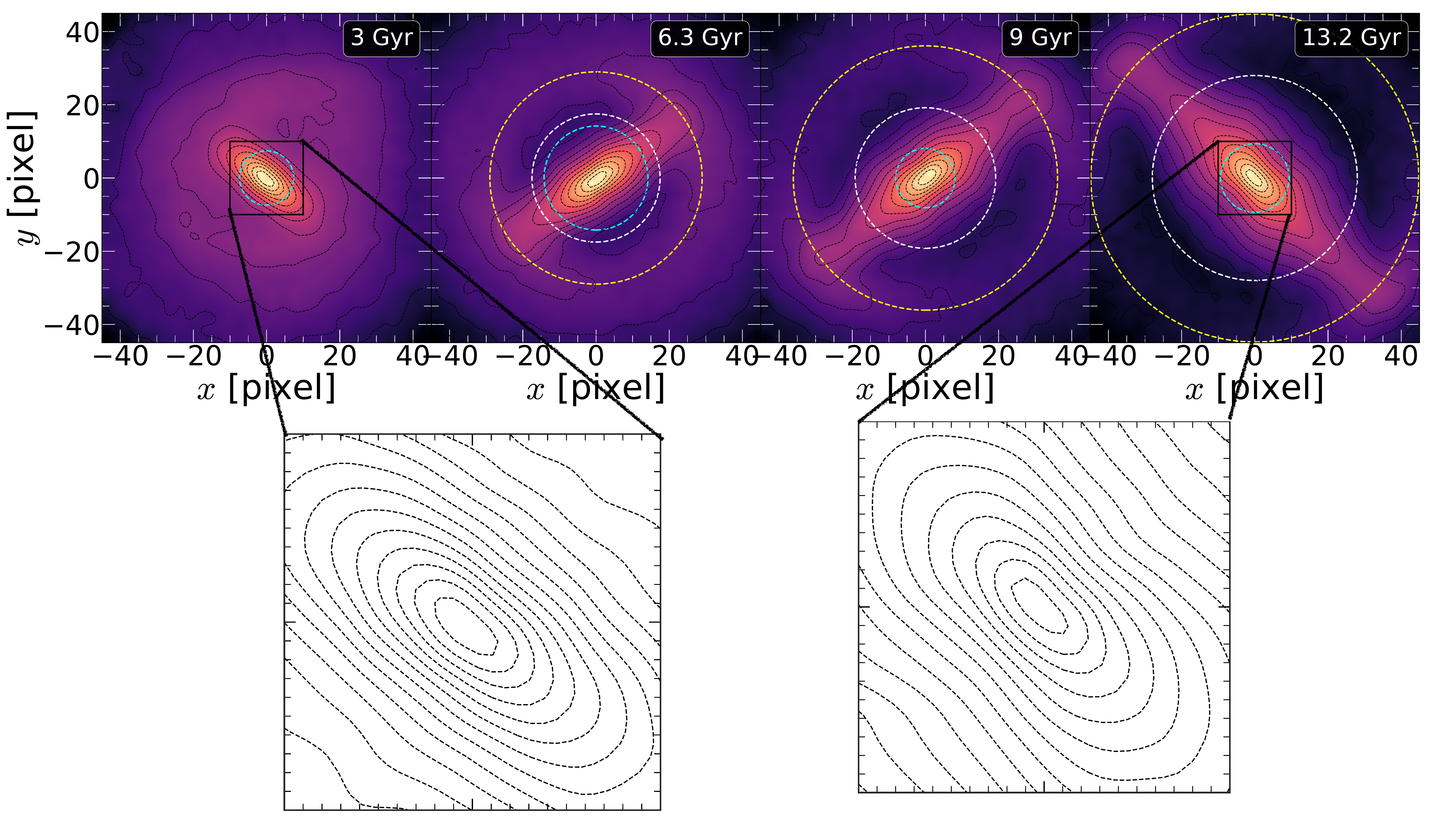}
\includegraphics[width=0.8\linewidth]{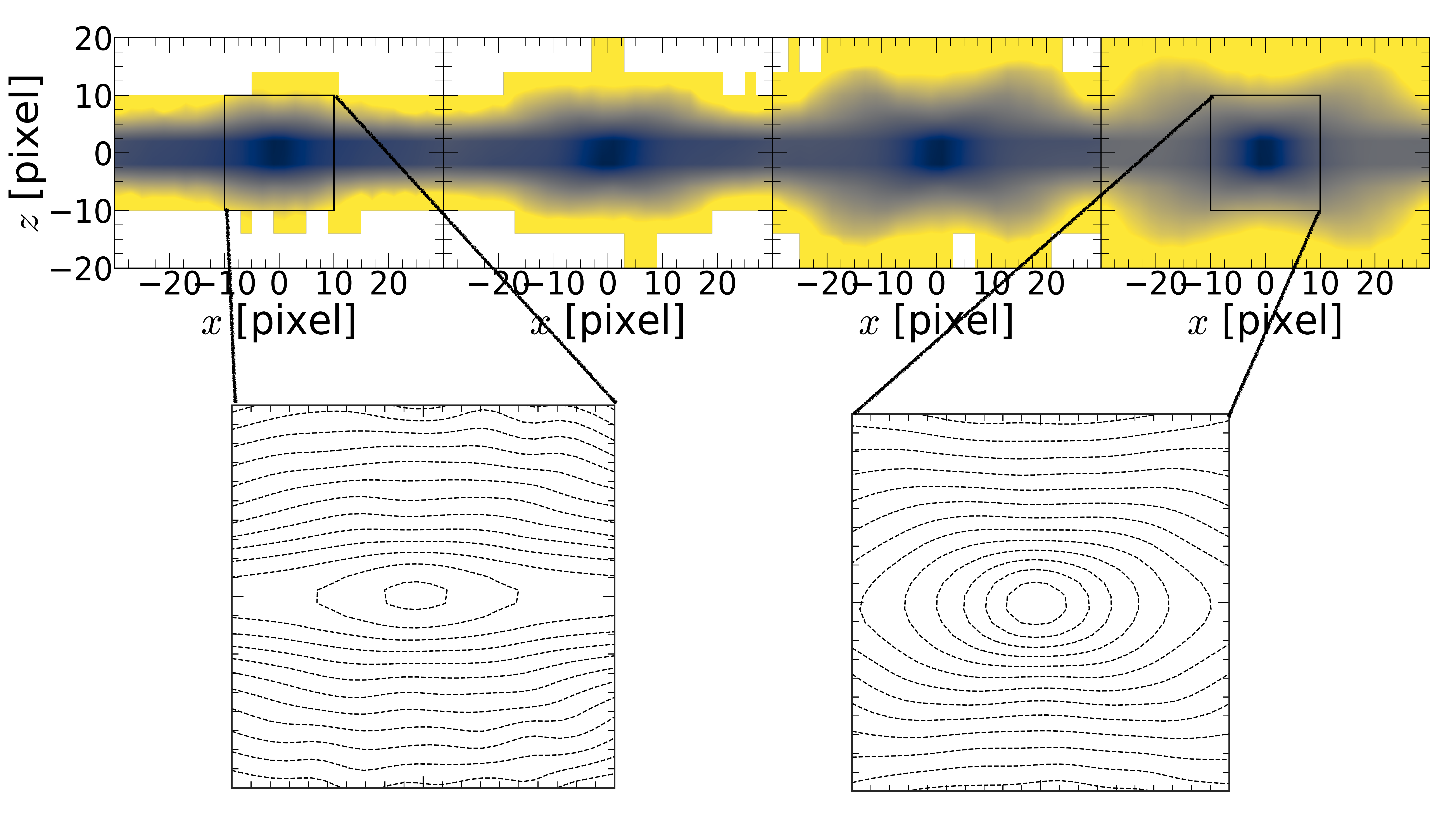}
\caption{{\it Top panels} show the face-on density distributions at four different epochs for the 3-D peanut model, depicting the scenario of transformation from a single-structured to a multiple sub-structured system in the bar region. {\it Bottom panels} show the corresponding edge-on density distributions at same four different epochs. The rectangular boxes (solid black lines) delineate the central $20\times 20 \pixel$ region (in each projection). The contours of these regions are further shown below each of these corresponding panels. The circles shown in the {\it top panels} denote the spatial locations of the peaks in $m=2$ Fourier mode (for details see text). Here, 1 pixel = $0.23 \kpc$.}
\label{fig:densitycollage_bpmodel}
\end{figure*}

\citet{MerrifieldKuijken1995} derived a slightly modified form of Eq.~\ref{eq:tw_ori}, and the pattern speed ($\Omega_{\rm p}$) of the bar is measured using the equation

\begin{equation}
\Omega_{\rm p} \sin i= \frac{\left<V \right>}{\left< X \right>}\,,
\label{eq:tw_actual}
\end{equation}
\noindent where,
\begin{equation}
\left<V \right> = \frac{\int_{- \infty}^{\infty} \Sigma(X, Y) V_{los} (X, Y) dX}{\int_{- \infty}^{\infty} \Sigma(X, Y) dX}; \vspace{0.2cm} 
\left< X \right> = \frac{\int_{- \infty}^{\infty} \Sigma(X, Y) X dX}{\int_{- \infty}^{\infty} \Sigma(X, Y) dX}\,.
\end{equation}
The integrals in Eq.~\ref{eq:tw_actual} are evaluated along a slit aligned with the disc position angle (PA). In the past, the TW method has been used extensively to measure the pattern speeds of bars in external disc galaxies \citep[e.g., see][]{MerrifieldKuijken1995,Gressenetal1999,Gressenetal2003,vpdetal2004,Fathietal2009,Aguerrietal2015,Cuomoetal2019,Guoetal2019,PatraandJog2019,Williamsetal2021}. However, it is associated with various uncertainty sources affecting the pattern speed measurements \citep[for details see e.g.,][]{vpd2003,Zouetal2019,Garmaetal2020}. Also, the importance of a  proper choice of a tracer (obeying the continuity equation) in measuring the bar pattern speed is further discussed in a recent study by \citet{Williamsetal2021}.

\section{Formation of multiple sub-structures in the 3-D peanut model}
\label{sec:multiple_structure}

Fig.~\ref{fig:densitycollage_bpmodel} shows the density distribution of stellar particles in the face-on and edge-on projections at four different times for our 3-D peanut model. At initial epochs (e.g., $t = 3 \Gyr$), the model harbours a prominent central bar, but the face-on peanut structure is absent. After $t= 6 \Gyr$ or so, the model starts developing a face-on peanut structure. This structure continues to grow, and by the end of the simulation run ($t= 13.2 \Gyr$), the model hosts a strong face-on peanut structure; see \cite{PatsisKatsanikas2014} for a detailed understanding of the orbital configuration in a similar situation where boxy/peanut structure is seen in the equatorial plane of a bar. A closer look at the central region revealed the change of shape of the density contours in the face-on configuration. Initially, e.g., at $t = 3 \Gyr$, the contours in the central region (delineating the strong bar) were very elongated elliptical. However, at the end of the simulation run ($t=13.2 \Gyr$) when the model hosts a strong face-on peanut, the contours in the central region have become rounder and bear a close resemblance with a `dumb-bell'-like configuration. A similar change of appearance of the density contours in the central region is evident from the edge-on configuration. At $t=13.2 \Gyr$, the density contours in the central region become more puffed-up in the vertical direction when compared with density contours at the initial time-steps (e.g. $t = 3 \Gyr$, see Fig.~\ref{fig:densitycollage_bpmodel}).

\begin{figure}
    \includegraphics[width=\linewidth]{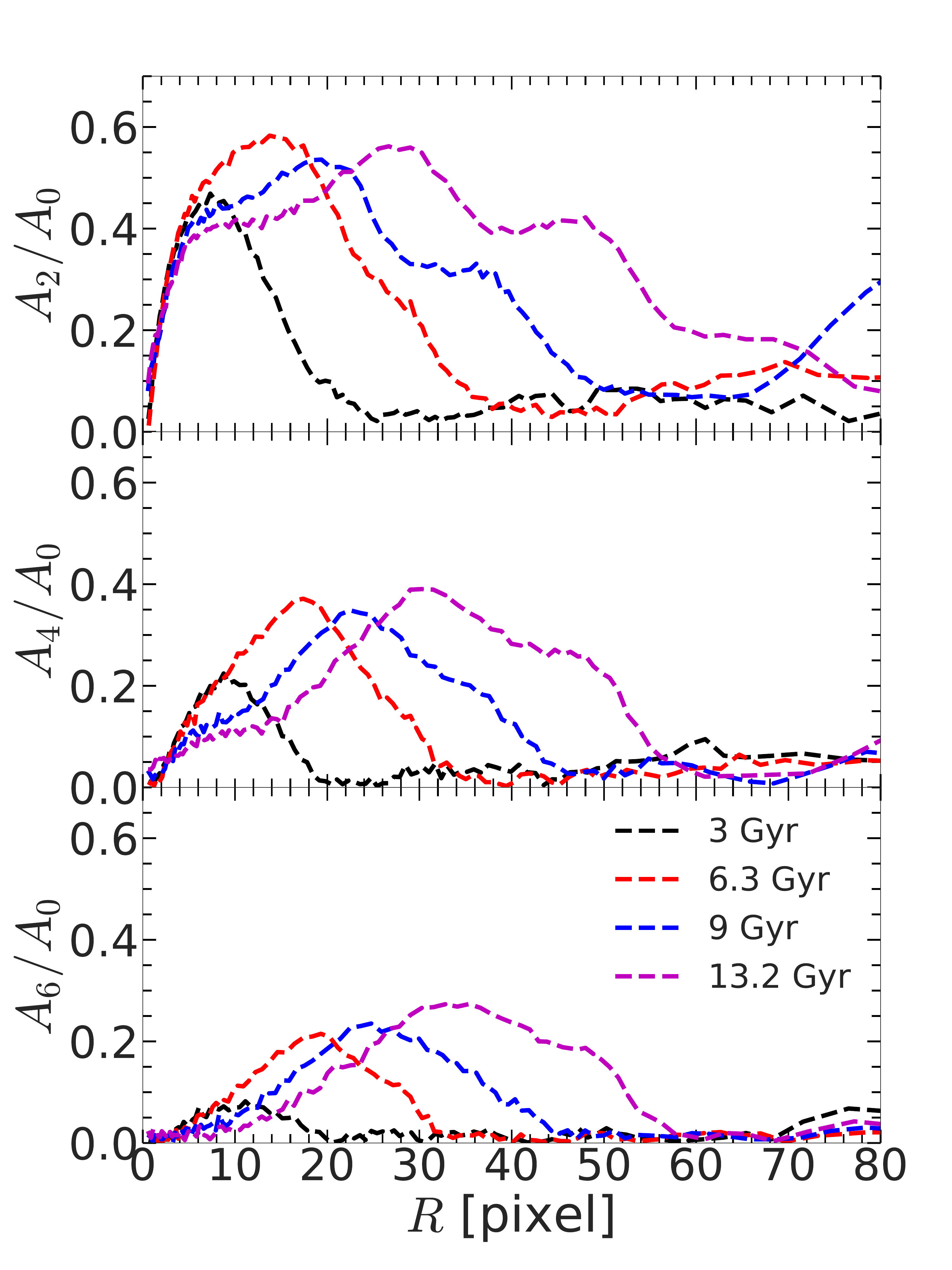}
\caption{Radial profiles of $m=2, 4, 6$ Fourier coefficients, calculated at four different epochs are shown for our 3-D peanut model. At later epoch, the radial profiles of the $m=2$ Fourier coefficient display multiple distinct peaks. Here, 1 pixel = $0.23 \kpc$.}
\label{fig:densfourier}
\end{figure}

\begin{figure}
\includegraphics[width=\linewidth]{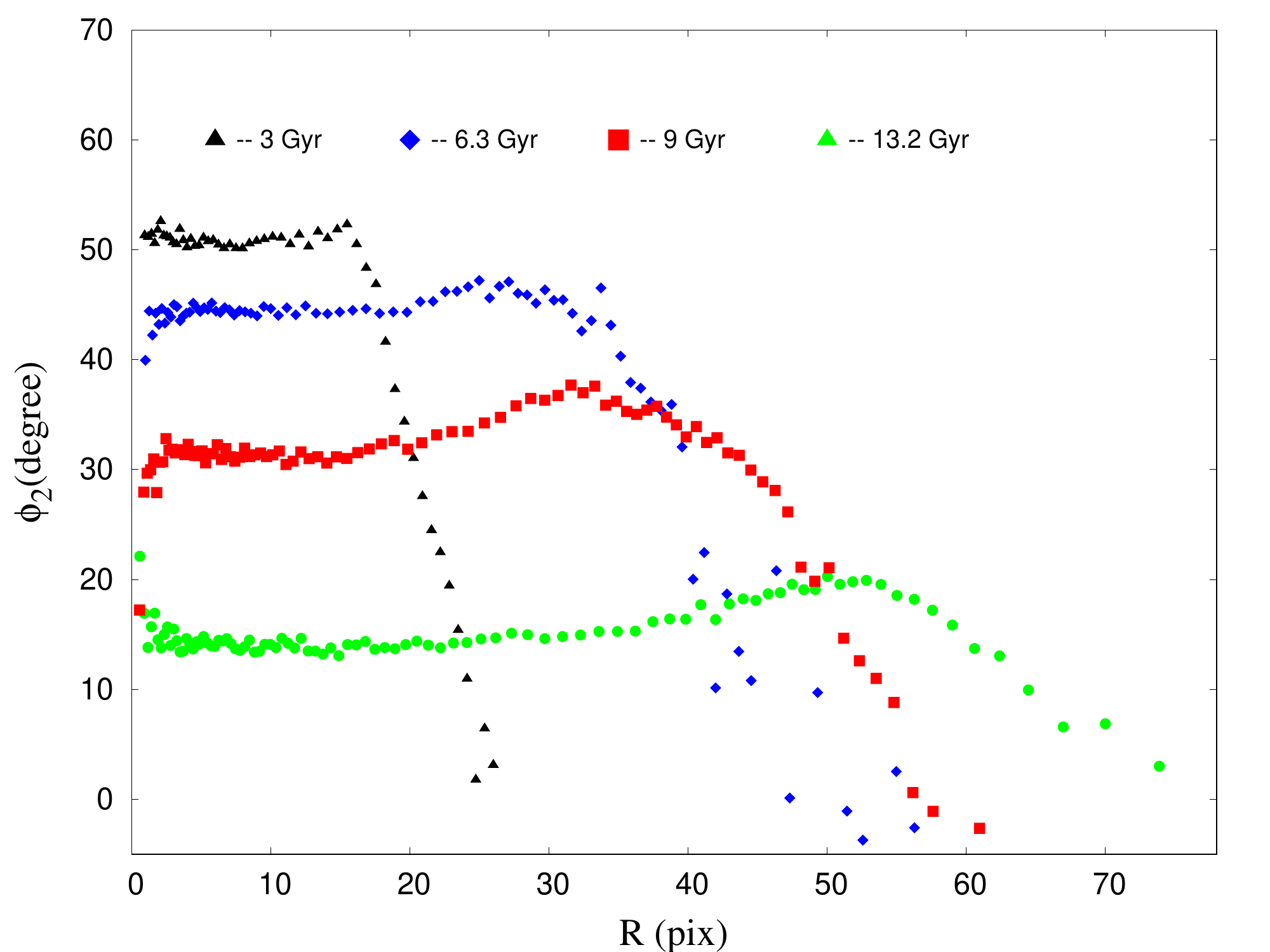}
\caption{Radial variation of the phase angle ($\phi_2$) corresponding to the $m=2$ Fourier component, calculated from the intrinsic particle distribution are shown at four different epochs. Here, 1 pixel = $0.23 \kpc$}
\label{fig:phi2}
\end{figure}

To investigate further, we calculate the radial variations of different azimuthal Fourier components of the mass distribution in the disc. A strong bar induces an $m=2$ Fourier mode as well as an $m=4$ Fourier mode (albeit weaker than the $m=2$ mode) whereas a b/p structure can be represented by an $m=6$ Fourier mode \citep{CiamburGraham2016,Sahaetal2018,Ciamburetal2020}. Fig.~\ref{fig:densfourier} shows the resulting radial variations of the $m=2, 4, 6$ Fourier harmonics, calculated at four different time-steps for the 3-D peanut model. As seen clearly, the radial profiles for the $m=2$ Fourier coefficient at later time-steps display three distinct peaks as opposed to a singly-peaked radial profile at the initial time-steps (e.g., at $t = 3 \Gyr$). The intermediate peak in the radial profiles of the $m=2$ Fourier coefficient started emerging around 6 Gyr. The appearance of the third (outer) peak follows shortly after the appearance of the intermediate peak in the radial profiles of the $m=2$ Fourier coefficient. After $t = 6.1 \Gyr$, the radial profiles of the $m=2$ Fourier coefficient display three distinct peaks, similar to what is shown at $t = 13.2 \Gyr$. Also, the radial profiles of the $m=6$ Fourier coefficient display a higher value at later times when compared to that for the initial epochs (e.g., say $t = 3 \Gyr$);  thereby denoting the presence of a strong peanut-shaped structure in the model at later time-steps. Furthermore, the peaks of the radial profiles for the $m=4$ and the $m=6$ Fourier coefficients shift towards outer radii at later time-steps. 
Besides the amplitude of the $m=2$ Fourier component, the phase angle ($\phi_{2}$) bears essential information on the dynamical structure of the bar. In Fig.~\ref{fig:phi2} shows the radial variation of the phase angle, $\phi_{2}$, at the four different times mentioned above. One of the defining characteristics of a bar is that the phase angle should remain constant over a range of radii; in other words, $d\phi_{2}/dR = 0$ within the bar region. This is roughly true at $t=3$~Gyr when the bar is small and has not buckled yet. At later times, the bar grows in size and enters the buckling phase due to the excitation of bar orbits at the vertical resonances \citep{,Combesetal1990,Athanassoula2003,Martinez-Valpuestaetal2006,Collier2020,SellwoodGerhard2020}. The phase angle starts displaying characteristic radial variation, especially in the outer parts containing the second and third peaks in the $m=2$ Fourier component. Such radial variation within the bar region may indicate that the whole bar may not be rotating rigidly or coherently. However, since the variation is only about a few degrees ($\sim 5^{\circ}$ between the inner and outer bar), the rigidity of the bar might remain intact for all practical purposes.

F
\begin{figure}
\includegraphics[width=\linewidth]{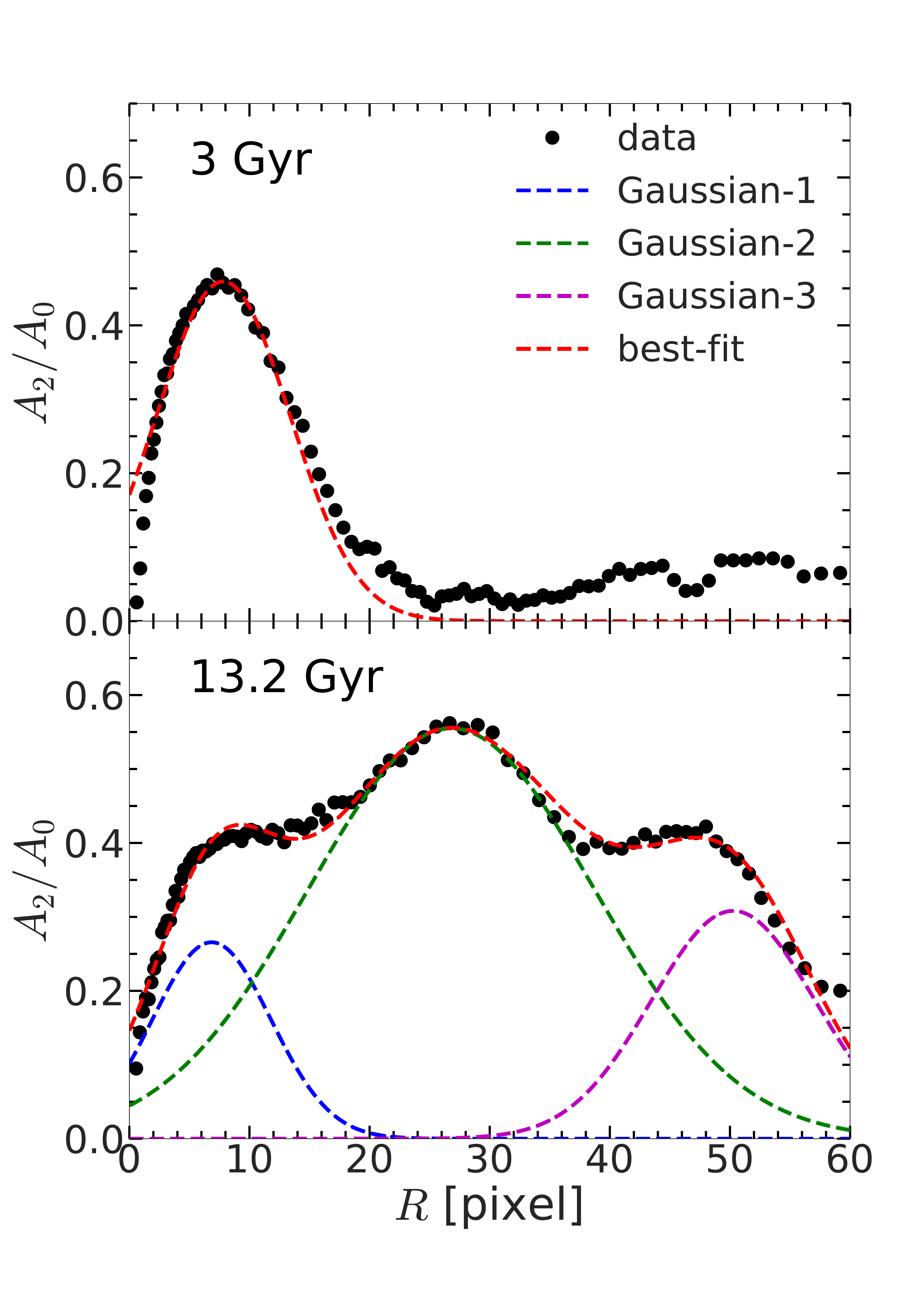}
\caption{Radial profiles of the $m=2$ Fourier coefficient at two different time-steps are shown. In each case, the peak(s) in the radial profile is modelled with a Gaussian profile (for details see text). The presence of three well-defined peaks at $t = 13.2 \Gyr$ demonstrates the co-existence of multiple sub-structures in the bar region.}
\label{fig:densfourier_comp}
\end{figure}

\begin{figure}
\includegraphics[width=\linewidth]{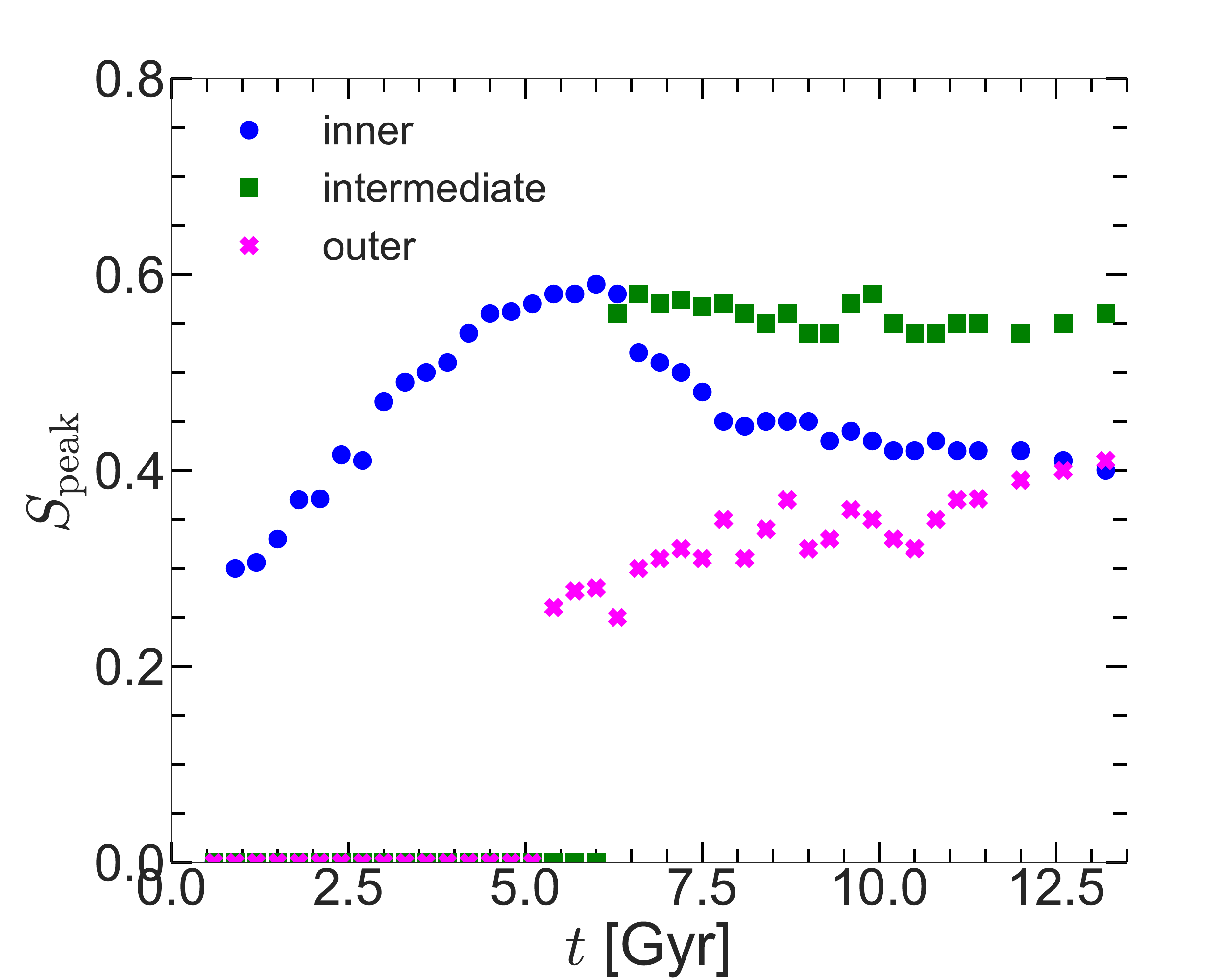}
\caption{Temporal evolution of the strengths (calculated by using Eq.~\ref{eq:peak_strength}) of the three distinct peaks, present in the radial profiles of $m=2$ Fourier harmonics,  are shown.}
\label{fig:temp_barpeak}
\end{figure}

To characterise further the trend seen in the number of peaks of the radial profiles for the $m=2$ Fourier coefficient, we model each of these peaks with a Gaussian profile. This is shown in Fig.~\ref{fig:densfourier_comp}. At initial times (e.g., $t = 3 \Gyr$), there is a single well-defined peak in the radial profile of $m=2$ Fourier coefficient, which is modelled with a single Gaussian profile (FWHM $\sim$ 13.1 pixel). However, at later times (e.g., $t = 13.2 \Gyr$), three separate Gaussian profiles are used to model each of these three well-defined peaks present in the radial profile $m=2$ Fourier coefficient. The second/middle peak shows the highest peak value ($\sim 0.56$) and has the maximum width (FWHM $\sim$ 28.1 pixel). On the other hand, the inner/first peak displays the lowest width (FWHM $\sim$ 11.6 pixel) and has the peak value ($\sim 0.4$), comparable to the third/outer peak (peak value $\sim 0.42$; FWHM = 16 pixel). We checked that the intermediate/second peak location approximately matches with the location of the ansae. In contrast, the location of the outer/third peak approximately matches with the `handle' of the long bar present in our 3-D peanut model \citep[for details see][]{Sahaetal2018}. 

At this point, it is worth investigating how the strengths of these three peaks in the radial profile of $m=2$ Fourier coefficient evolve with time. To do that, we first define the strength of the peak ($S_{\rm peak}$) as the highest value of the $A_2/A_0$ coefficient corresponding to the peak, i. e.,

\begin{equation} 
S_{\rm peak} = (A_2/A_0)_{\rm max}\,.
\label{eq:peak_strength}
\end{equation}

\noindent Fig.~\ref{fig:temp_barpeak} shows the resulting temporal evolution of the strength of the peaks. Initially, the inner/first peak grows with time, indicating the growth of the bar; however, after $\sim 6.6 \Gyr$ it starts to weaken. The epochs for the growth of the second (intermediate) and the third (outer) peak approximately matches with the epoch of the weakening of the first (inner) peak. At the end ($t = 13.2 \Gyr$), the intermediate peak remains strongest, showing the highest value of the $m=2$ Fourier coefficient ($A_2/A_0$). The presence of three well-defined peaks in the radial profile $m=2$ Fourier harmonics at later times demonstrates the existence of multiple sub-structures encompassing the whole bar region, as opposed to the scenario of a single bar structure seen at earlier time-steps in our 3-D peanut model. 

\begin{figure*}
\includegraphics[width=0.9\linewidth]{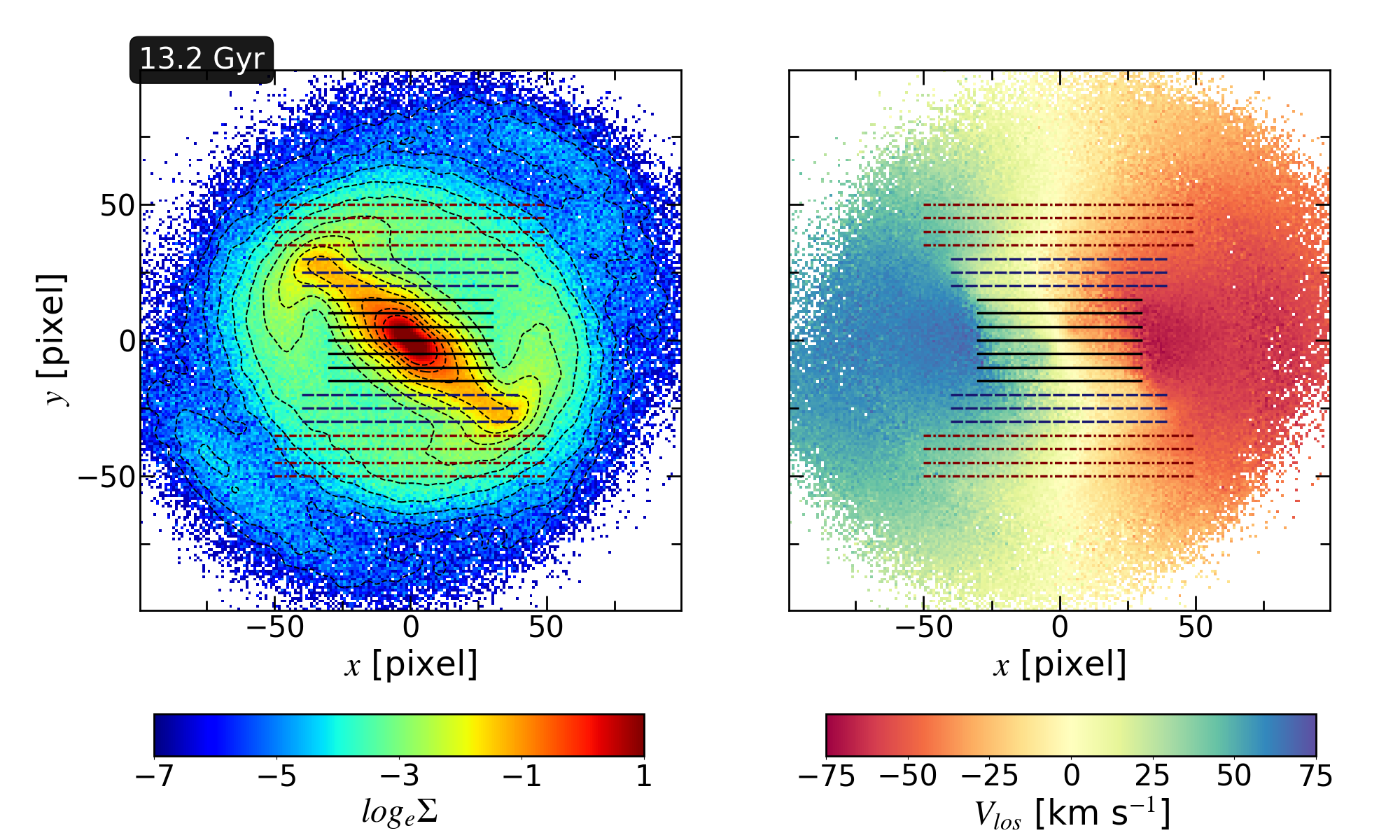}
\caption{Line-of-sight density and velocity distribution (angle of inclination $i=30\degrees$) of the 3-D peanut model, calculated at $t=13.2 \Gyr$. The individual colour bars are indicated in each sub-panel. The horizontal lines denote the positions of the (pseudo) slits, covering different bar regions, which are used to measure the bar pattern speed via the TW method (for details, see text). The half-slit lengths ($L_0$) for the inner, intermediate, and outer regions are 30, 40, and 50 pixels, respectively. Here, 1 pixel = $0.23 \kpc$.}
\label{fig:sim_2}
\end{figure*}

\section{Pattern speed measurement in the 3-D peanut model}
\label{sec:pattern_speed_measurement}
%

The previous section demonstrated that the 3-D peanut model considered here develops multiple sub-structures in the bar region at later epochs. However, the question remains whether these multiple sub-structures in the long bar region rotate in unison with a single pattern speed or with their characteristic pattern speeds that differ from one another. To address this, we measure the pattern speeds in the bar region using the TW method described in section~\ref{sec:tw_method}. In order to apply the TW method, we first construct the line-of-sight density and velocity distribution (at $i=30\degrees$) for our model, calculated from the intrinsic six-dimensional position-velocity of the stellar particles. Fig.~\ref{fig:sim_2} shows such line-of-sight density and velocity distributions calculated at $t = 13.2 \Gyr$. A prominent stellar bar, as well as the signature of a face-on peanut structure, is seen in the density distribution \citep[for details see][]{Sahaetal2018}.

Next, we measure the pattern speed of the bar present in the 3-D peanut model. We caution the reader that co-existing multiple sub-structures in the bar region (as shown in section~\ref{sec:multiple_structure}) might rotate with their respective pattern speeds, different from one another. In such a scenario, applying the TW method for the whole bar region would be misleading, as different patterns (if any) will interfere with each other, causing an erroneous measurement of the pattern speed ($\Omega_{\rm p}$). To avoid such a scenario, we first divide the entire bar region into three parts so that each sub-region contains the corresponding peak in the radial profile of the $m=2$ Fourier mode. We refer to them as inner, intermediate, and outer regions in the rest of this paper. In other words, the inner region always contains the inner/first peak in the radial profile of the $m=2$ Fourier coefficient, and so on. Then, the TW method is applied in each of these three sub-regions \textit{separately} to measure the corresponding pattern speeds. We use different slit lengths for these three regions while measuring the pattern speed. The values of the half-slit length ($L_0$) for the inner, intermediate, and outer regions are 30, 40, and 50 pixels, respectively.

\subsection{Error estimation in pattern speed measurements}
%
 Here, we first estimate/quantify the uncertainty sources, for example, the disc kinematic position angle (hereafter, PA) measurement, disc inclination angle, disc centring, length and width of the slits used \citep[for details see][]{vpd2003,Zouetal2019,Garmaetal2020}, in measuring the pattern speeds using the TW method. The error in determining the disc PA can induce a significant error in the pattern speed measurement, as shown previously by \citet{vpd2003,Garmaetal2020}. We checked that the disc kinematic PA remains $\sim 0 \degrees$ for all the time-steps used (later) for the pattern speed measurement, thereby justifying our choice of horizontal slits while employing the TW method. Here, we stress that, for a simulated model with known six-dimensional phase-space information for all stellar particles, we can measure the disc kinematic PA, disc centre, and inclination angle quite robustly, unlike in observational studies. Thus, uncertainties coming from these sources are negligible in this case.

\begin{figure*}
\includegraphics[width=0.8\linewidth]{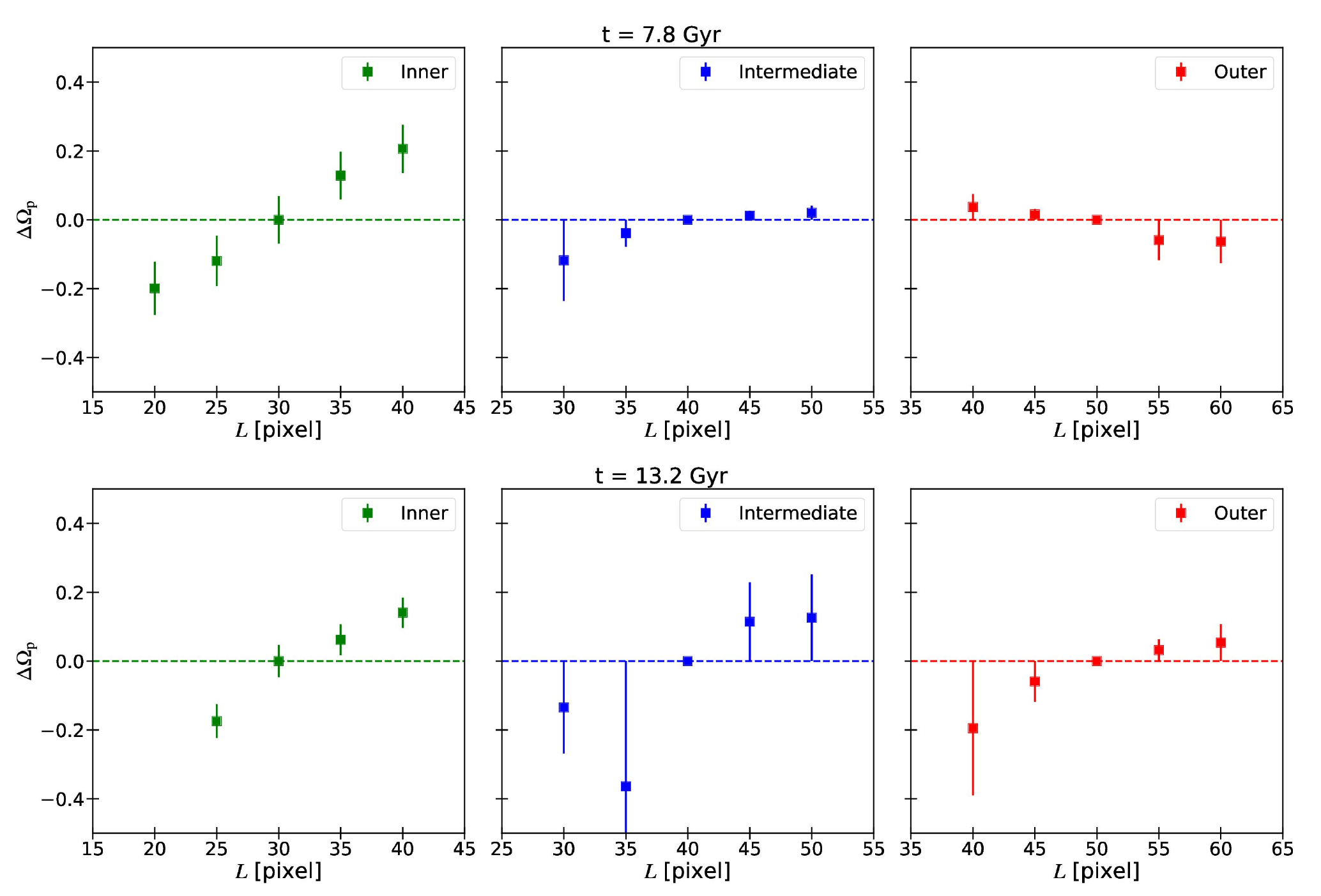}
\caption{Variation of the fractional error ($\Delta \Omega_{\rm p}$) in pattern speed measurement (see Eq.~\ref{eq:err1}) with varying half-slit length ($L$) is shown at $t = 7.8 \Gyr$ and $t = 13.2 \Gyr$ for our 3-D peanut model. The reference values of the half-slit length ($L_0$) for the three regions are $30 \pixel$, $40 \pixel$ and $50 \pixel$, respectively. The dashed lines in each of these sub-panels correspond to null fractional error in pattern speed measurement.}
\label{fig:patternspeed_slitlengthvary}
\end{figure*}

Next, we estimate the error in pattern speed measurement due to the variation in the slit length. We point out that the correct choice of slit length is vital, as we are trying to measure the pattern speeds at different spatial locations in the bar region. Ideally, a slit should cover a particular region (e.g., inner, intermediate or outer region) to yield an accurate measurement of pattern speed for that region. First, we varied the half-slit lengths ($L$) by up to  30-40 per cent about their assumed values ($L_0$, see section~\ref{sec:pattern_speed_measurement} for details), and measured the fractional error in the pattern speed measurements. The fractional error ($\Delta \Omega_{\rm p}$) in the pattern speed measurement is calculated in the following way.

\begin{equation}
\Delta \Omega_{\rm p} (L) = \frac{\Omega_{\rm p}(L) - \Omega_{\rm p} (L_0)}{\Omega_{\rm p} (L_0)}\,,
\label{eq:err1}
\end{equation}
\noindent where $L$ is the half-slit length, and $L_0$ is the initially chosen half-slit length for a particular region (see section~\ref{sec:pattern_speed_measurement} for actual chosen values of $L_0$). The resulting variation of measured pattern speed values with varying slit-length are shown in Fig.~\ref{fig:patternspeed_slitlengthvary} for $t = 7.8 \Gyr$ and  $t = 13.2 \Gyr$. A saturation in the fractional error in pattern speed values with varying slit-length is a reasonably good indicator for an appropriate slit length \citep[for more details see, e.g.,][]{Zouetal2019,Garmaetal2020}. We found that, for varying slit lengths, the fractional error in pattern speed measurements for the three regions considered here, is below 20 per cent for most of the cases.
\begin{figure*}
\includegraphics[width=0.8\linewidth]{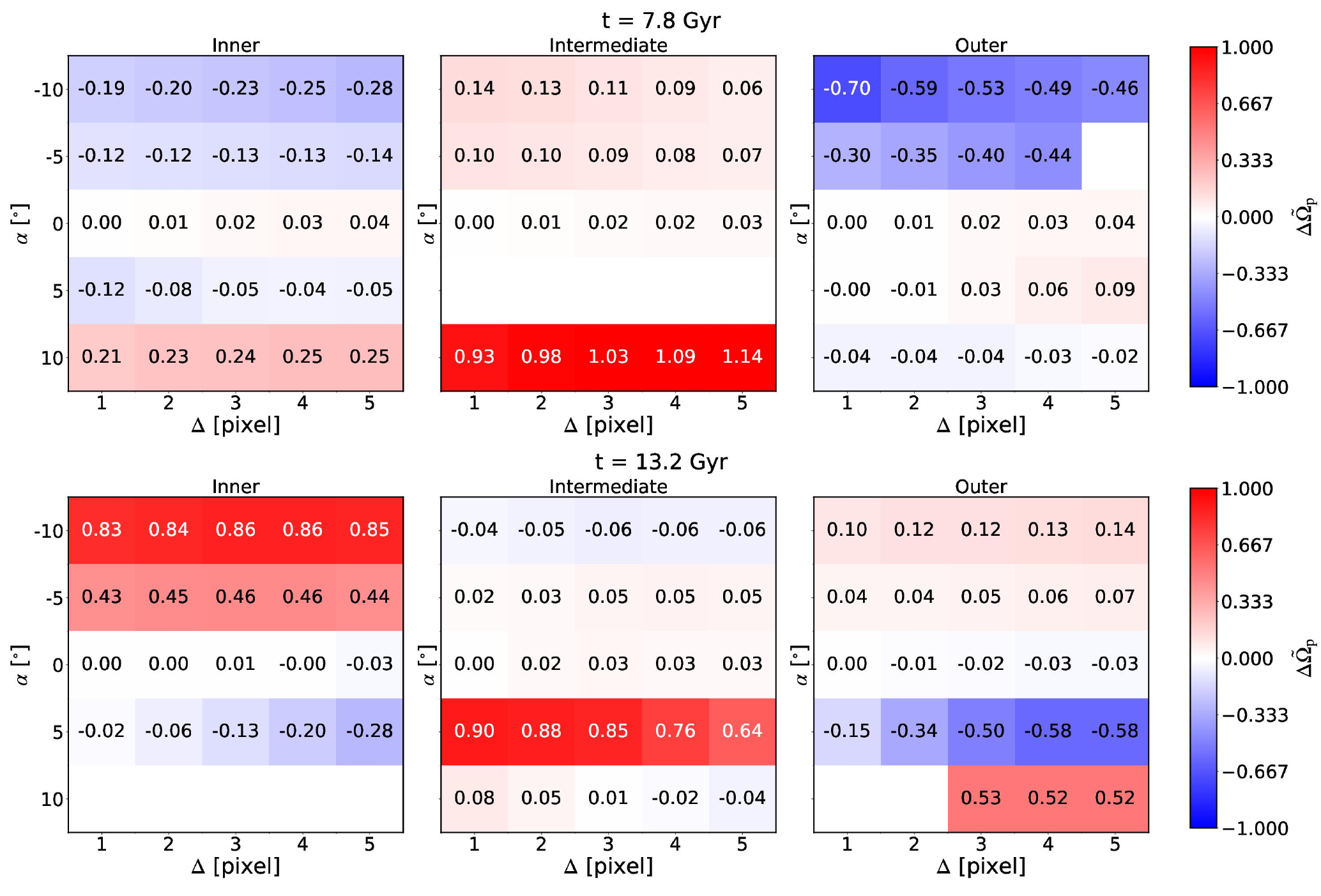}
\caption{Distribution of the fractional error ($\Delta \tilde \Omega_{\rm p}$) in pattern speed measurement (Eq.~\ref{eq:err2}) with varying slit width ($\Delta$) and the disc kinematic PA ($\alpha$) is shown at $t = 7.8 \Gyr$ and $t = 13.2 \Gyr$ for our 3-D peanut model. The reference values for $\Delta_0$ and $\alpha_0$ are $1 \pixel$ and $0 \degrees$. Values with absolute errors greater than $2 \,\text{km s}^{-1}\text{ kpc}^{-1}$ are omitted.}
\label{fig:patternspeed_slitheightinclvary}
\end{figure*}

Lastly, we estimate the systematic errors induced in the pattern speed measurement due to the joint variation in the slit width ($\Delta$) and the disc kinematic PA ($\alpha$). The fractional error due to the joint variation of these two quantities is calculated in the following way:
\begin{equation}
\Delta \tilde \Omega_{\rm p} (\Delta, \alpha) = \frac{\Omega_{\rm p} (\Delta, \alpha) - \Omega_{\rm p}(\Delta_0, \alpha_0)}{\Omega_{\rm p}(\Delta_0, \alpha_0)}\,,
\label{eq:err2}
\end{equation}
\noindent where $\Delta$ and $\alpha$ denote the slit width and the disc kinematic PA respectively, and the subscript `0' denotes the initially-chosen values of the slit width and the disc kinematic PA. Fig.~\ref{fig:patternspeed_slitheightinclvary} shows the resulting distribution of the fractional error in the pattern speed measurement as a function of varying slit width and the disc kinematic PA, at $t = 7.8 \Gyr$ and  $t = 13.2 \Gyr$. For a fixed disc kinematic PA, the error in the pattern speed measurements due to variations of the slit width is small ($\sim 5-15$ per cent), in concordance with the previous studies \citep[e.g., see][]{Cuomoetal2019,Guoetal2019,Garmaetal2020}. The larger contribution in error comes from the variation of the disc kinematic PA; particularly when the slit position coincides with the bar orientation (e.g, the cases with most blue and red in Fig.~\ref{fig:patternspeed_slitheightinclvary}). Fortunately, as mentioned before, the measurement of disc PA in a simulated model (like ours) is quite robust.

\subsection{Pattern speed measurement and its temporal evolution}
\label{sec:temp_evolution_pattern_speed}

Finally, we measure the pattern speeds of the three regions in our model by employing the TW method. We caution that we have measured the pattern speed values at only those epochs when the bar is oriented neither along the galaxy's major axis nor along the minor axis, thereby avoiding unreliable pattern speed measurements obtained by the TW method \citep[for details see, e.g.,][]{Cuomoetal2019,Garmaetal2020}. 

\begin{figure}
\includegraphics[width=1.\linewidth]{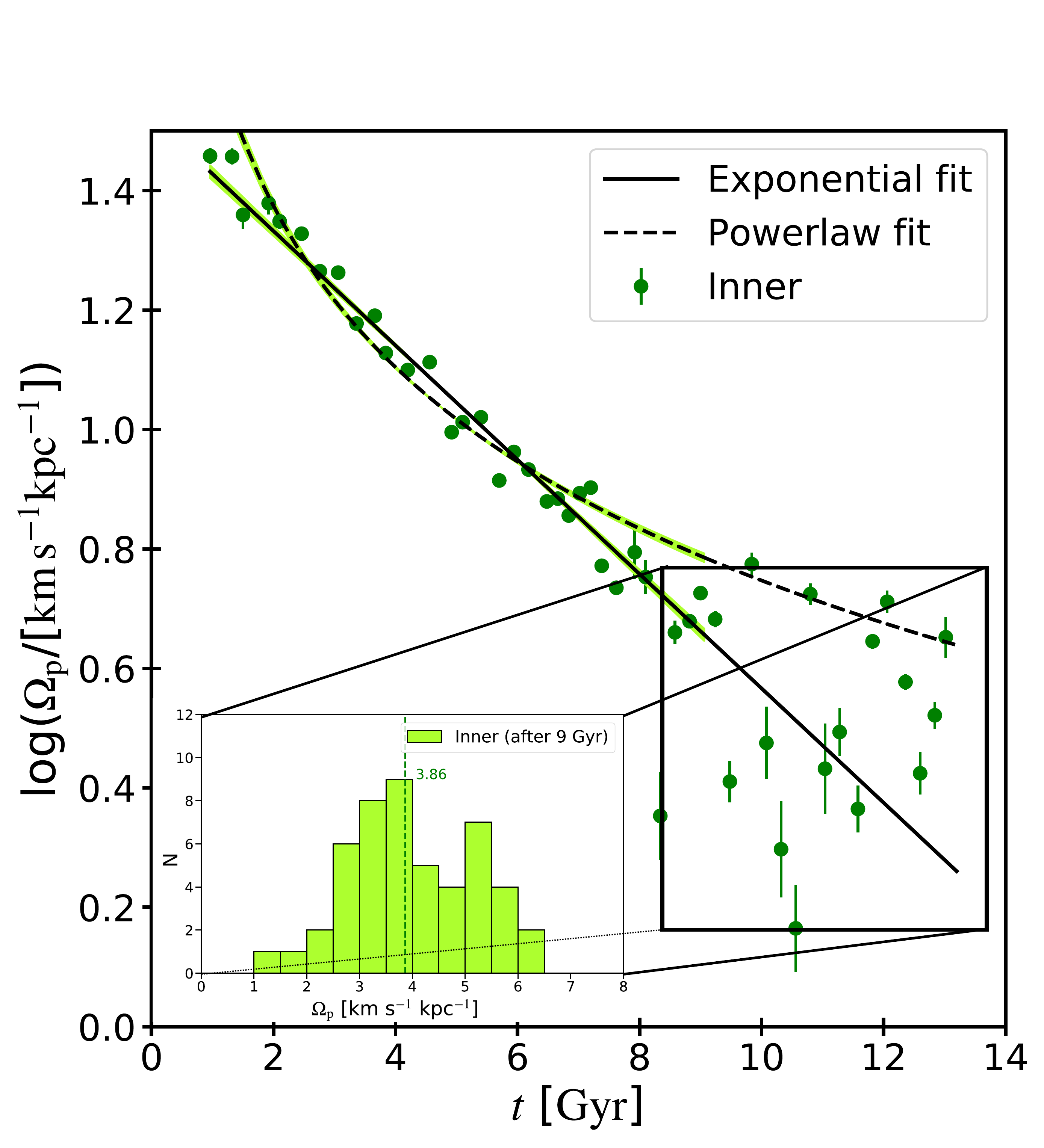}
\caption{Temporal evolution of the pattern speed (in log scale) of the inner region are shown till $t = 13.2 \Gyr$. An exponential profile and a power-law profile (of the forms as given in Eqs. \ref{eq:exp_fit} and \ref{eq:pow_fit}) are fitted to the temporal evolution, but the fitting is restricted to $t = 9 \Gyr$ (for details, see text). Based on the reduced $\chi^2$ values of the two fits, the exponential fit is judged to be better. Note that only every third (valid) point is plotted here to avoid overcrowding. The histogram, however, considers all (valid) points after $t = 9 \Gyr$.}
\label{fig:patternspeed_inner}
\end{figure}

First, we focus on the pattern speed measurements for the inner region, which contains the core of the peanut structure. The resulting temporal evolution of the pattern speed values ($\Omega_{\rm p}$) is shown in Fig.~\ref{fig:patternspeed_inner}. The pattern speed values decrease monotonically with time, which remains true up to $t = 9 \Gyr$. After $t=9 \Gyr$, the resulting pattern speed values are scattered and do not follow any specific trend as revealed by visual inspection. This could be because the values are close to zero, and the points are plotted in the logarithmic scale. To quantify further the rate of decrease in the pattern speed values measured in the inner region, we fit an exponential profile of the form 
\begin{equation}
\Omega_{\rm p} (t) = \Omega_{\rm p,0} e^{-\lambda t}\,,
\label{eq:exp_fit}
\end{equation}
\noindent where $\Omega_{\rm p,0}$ and $\lambda$ are constants. The exponential profile is a non-linear least square fit employing the Levenberg-Marquardt algorithm \citep{More1978}. Since the pattern speed values after $t= 9 \Gyr$ are not well-behaved, we restrict our fitting process till $9 \Gyr$. The resulting best-fit parameters are $\lambda = 0.220 \pm 0.006$ Gyr$^{-1}$ and $\Omega_{\rm p,0} = 33.4 \pm 1.0 \,\text{km s}^{-1}\text{ kpc}^{-1}$. We further checked that, when the $\Omega_{\rm p}$ values measured after $t = 9 \Gyr$ are included in the fitting process, the best-fit curve significantly deviates from measured values near the start of the simulation. For the sake of completeness, we tested whether the decay of the pattern speed values in the inner region can be well-represented by a power-law of the form 
\begin{equation}
\Omega_{\rm p} (t) = \Omega'_{\rm p,0} t^{-\mu}\,,
\label{eq:pow_fit}
\end{equation}
\noindent where $\Omega'_{\rm p,0}$ and $\mu$ are constants, and $t$ is in $\Gyr$. As before, we restrict up to $t = 9 \Gyr$, and fit the power-law function to the temporal variation of the pattern speed. The resulting best-fit parameters are $\mu = 0.89 \pm 0.03$ and $\Omega'_{\rm p,0} = 44.1 \pm 1.8 \,\text{km s}^{-1}\text{ kpc}^{-1}$. We find that the reduced chi-square value for the power-law fit ($\chi^2_{\rm pow} \sim 43.5$) is significantly higher than that obtained for the exponential fit ($\chi^2_{\rm exp} \sim 29.1$). Here, both reduced chi-square values are calculated with the same number of degrees of freedom. We conclude that an exponential profile describes the decrease in pattern speed better than a power-law profile. Hence, for the other two regions, only the exponential fit is considered.

\begin{figure*}[!b]
\includegraphics[width=\linewidth]{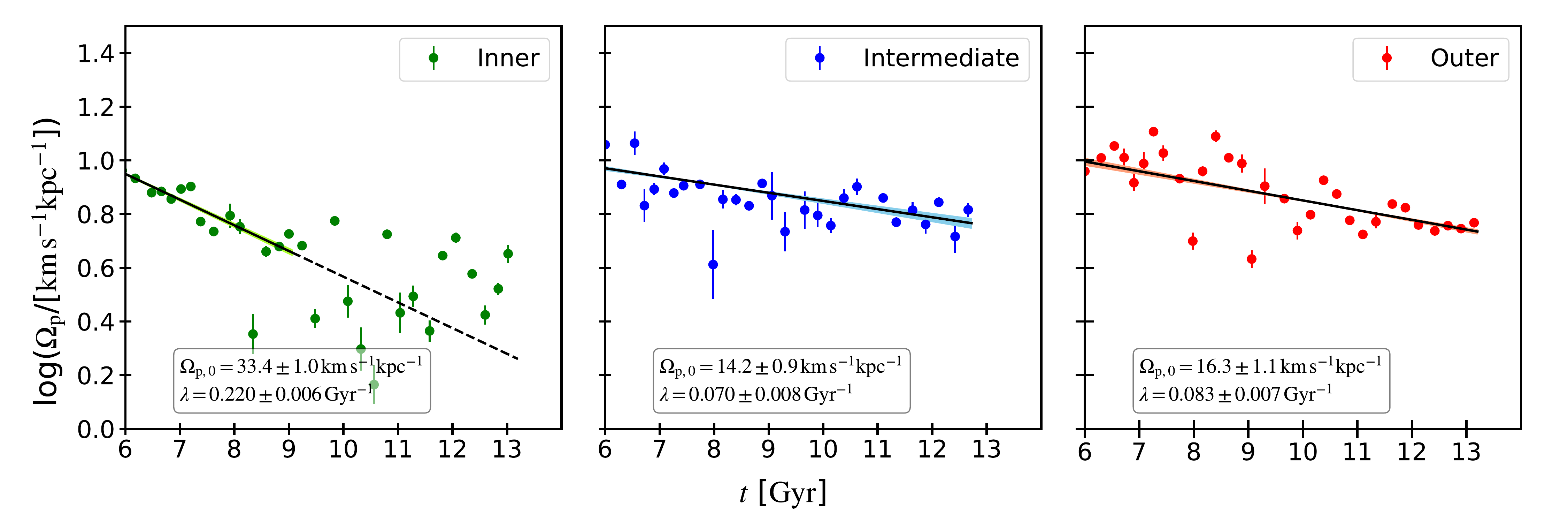}
\caption{Temporal evolution of the pattern speeds (in log scale), measured in three different regions, using the TW method (Eq.~\ref{eq:tw_actual}) for the 3-D peanut model. The black solid line in each sub-panel denotes the best-fit exponential function of the temporal variation of the pattern speed values. The corresponding best-fit values are mentioned in each sub-panel. Note that only every third (valid) point is plotted here to avoid overcrowding. }
\label{fig:patternspeed_bpmodel}
\end{figure*}

\begin{figure}
\includegraphics[width=\linewidth]{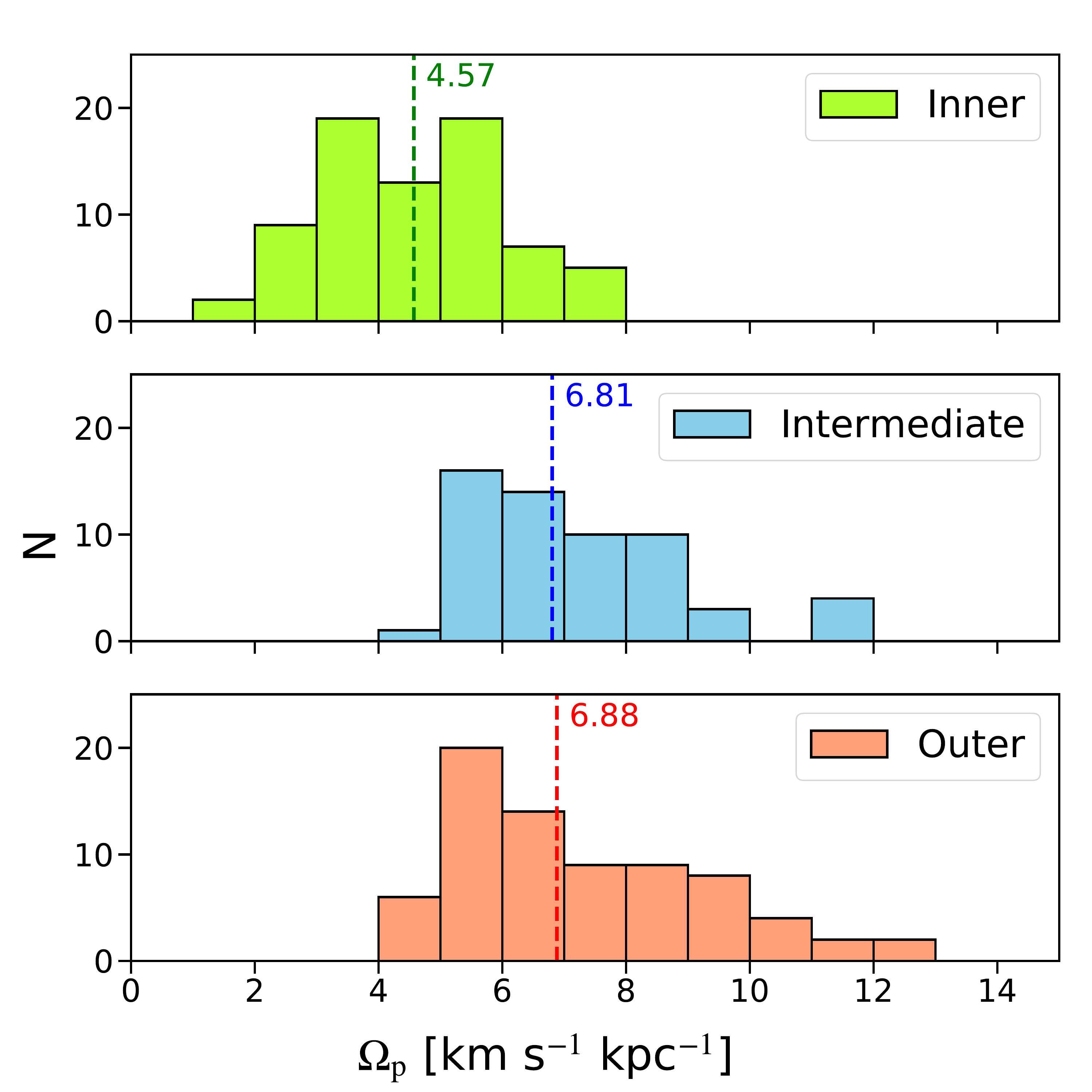}
\caption{The histogram of measured pattern speeds for the three regions, between $t = 7.2 \Gyr$ to $t = 13.2 \Gyr$, is shown for our 3-D peanut model. The dotted lines represent the median values (written above) of pattern speed in this time interval. The pattern speeds for the inner region are systematically lesser than those of the other two regions.}
\label{fig:patternspeed_histogram}
\end{figure}

Next, we study the temporal variation of the pattern speed values for the intermediate and the outer bar regions. This is shown in Fig.~\ref{fig:patternspeed_bpmodel}. For comparison, we kept the temporal variation of $\Omega_{\rm p}$ values for the inner region. The pattern speed values for the intermediate and the outer region decrease with time as well.  At the end of the simulation run ($t = 13.2 \Gyr$), the intermediate and the outer regions rotate with an approximately similar pattern speed. However, at the same epoch, the inner region rotates with a lower pattern speed.
The question remains whether the rate of decrease in the pattern speed values is similar or different for these three regions. We fit the same exponential profile, as given in Eq.~\ref{eq:exp_fit}, to the temporal variations of the $\Omega_{\rm p}$ values measured in the intermediate and the outer regions. The details of the best-fit values for these two regions are given below.

\begin{itemize}
 \item \textit{Intermediate region:} the best-fit parameters are $\lambda = 0.070 \pm 0.008$ Gyr$^{-1}$ and $\Omega_{\rm p,0} = 14.2 \pm 0.9 \,\text{km s}^{-1}\text{ kpc}^{-1}$.
       
    \item \textit{Outer region:}  the best-fit parameters are $\lambda = 0.083 \pm 0.007$ Gyr$^{-1}$ and $\Omega_{\rm p,0} = 16.3 \pm 1.1 \,\text{km s}^{-1}\text{ kpc}^{-1}$.    
\end{itemize}

The best-fit analysis demonstrates that the decrease in the pattern speed values for the inner region is indeed more rapid than for the intermediate and the outer regions. Lastly, Fig.~\ref{fig:patternspeed_histogram} shows the distribution of pattern speed values for the three different regions, calculated between $t = 7.2 \Gyr$ to $t = 13.2 \Gyr$, when our model displays multiple sub-structures encompassing the long bar region. The median values of the distributions further show that at later epochs, the inner region indeed rotates with a systematically lower pattern speed compared with the intermediate and the outer regions. This gives credence to our assumption of multiple pattern speeds in the bar.

\subsection{Decay time scale of the bar pattern speed}

We find that the parameter $\lambda$, which quantifies how fast the fall-off or the deceleration of the $\Omega_{\rm p}$ values, is almost three times higher for the inner region when compared with the intermediate and outer regions. To quantify further, following  \citet{Weinberg1985}, we define the decay time-scale of the bar ($\tau_{\rm bar}$) as $\tau_{\rm bar} = \frac{\Omega_{p}}{|{\dot \Omega_{p}}|}$. Now, for an exponential fall-off, the decay time-scale reduces to $\tau_{\rm bar} = 1/\lambda$. Using the best-fit parameters, as given in section~\ref{sec:temp_evolution_pattern_speed}, we calculate the decay time-scale ($\tau_{\rm bar}$) for the inner peanut region to be $4.5 \Gyr$ which is significantly shorter compared to the outer bar region (where $\tau_{\rm bar}$ is $>12 \Gyr$).  If this scenario holds for a long peanut bar, similar to what is considered here, then we might see a very slowly rotating inner peanut region in observations.

 Also, using the functional form as given in Eq.~\ref{eq:exp_fit}, we estimate the slowing down rate ($\dot \Omega_{\rm p}$) for three different regions considered here. At $t = 7 \Gyr$ when three structures co-exist, the estimated values of $\dot \Omega_{\rm p}$ are $-1.57 \kmskpcGyr$,  $-0.61 \kmskpcGyr$, and  $-0.74 \kmskpcGyr$ for the inner, intermediate, and the outer regions, respectively. At a later time, at $t = 9 \Gyr$, the estimated values of $\dot \Omega_{\rm p}$ are $-1 \kmskpcGyr$,  $-0.52 \kmskpcGyr$, and  $-0.63 \kmskpcGyr$ for the inner, intermediate, and the outer regions, respectively. 
 
Thus, to conclude, the long peanut bar present in our 3-D peanut model display three distinct peaks in the radial profiles of the $m=2$ Fourier mode at later epochs (after $t \sim 6.1 \Gyr$ or so). These multiple structures are shown to rotate with pattern speeds different from each other. Moreover, the rate of deceleration of the bar is different in these three regions. The pattern speed of the inner region falls relatively faster than for the intermediate and the outer regions. 

\section{Impact of multiple pattern speeds in the bar}
\label{sec:impact_patternspeed}

\begin{figure*}
\includegraphics[width=\linewidth]{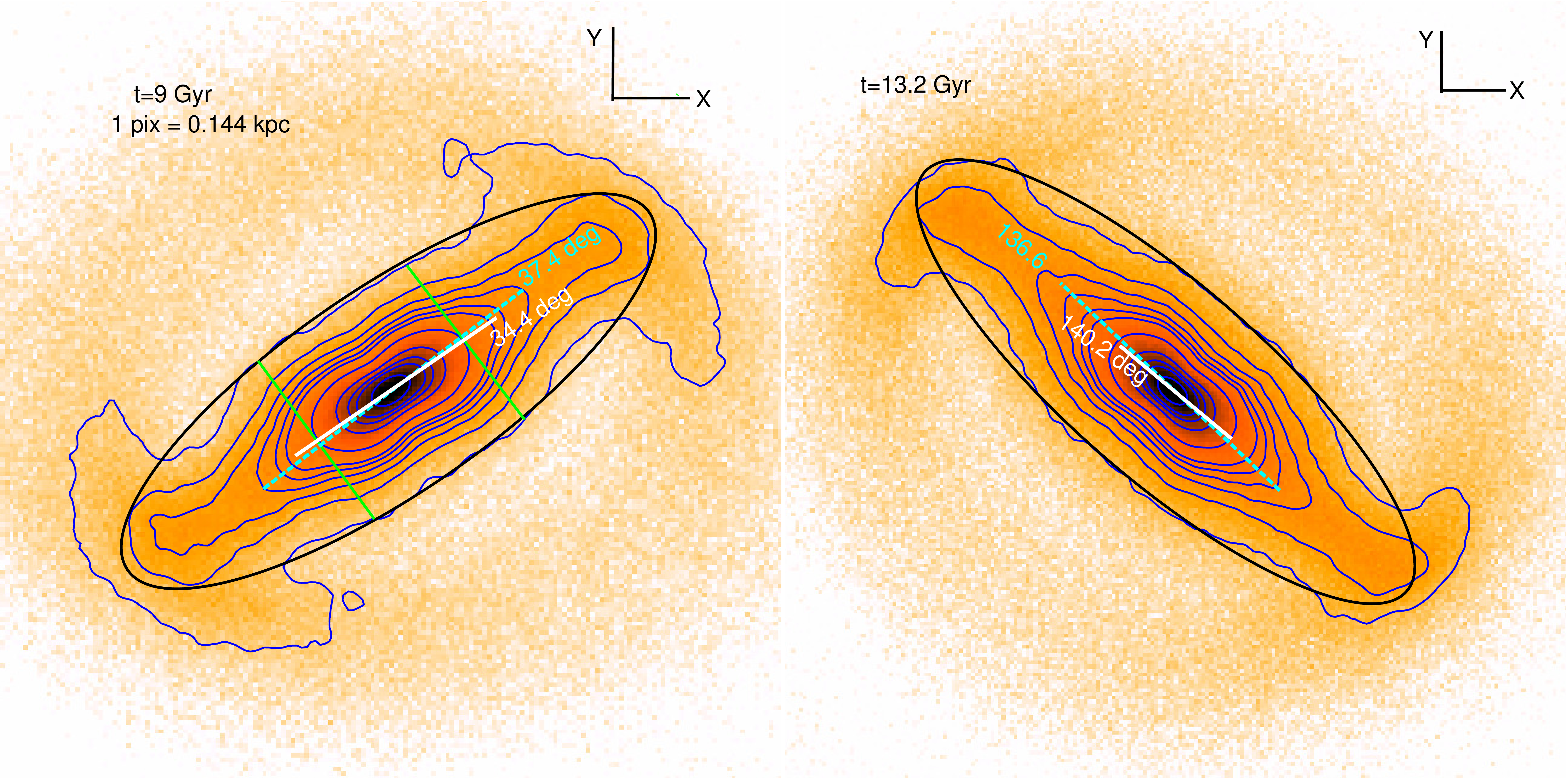}
\caption{Left panel: Density contours in the bar region. Contours are drawn at $7.,4.,3., 2.,1.,0.6,0.5,0.4,0.3,0.2,0.1$ from inner most to outer part of the bar in internal unit. The green lines parallel to the bar minor axis (at 9 Gyr) represent roughly the inner peanut region.}
\label{fig:densitycontour}
\end{figure*}

\begin{figure}
\includegraphics[width=\linewidth]{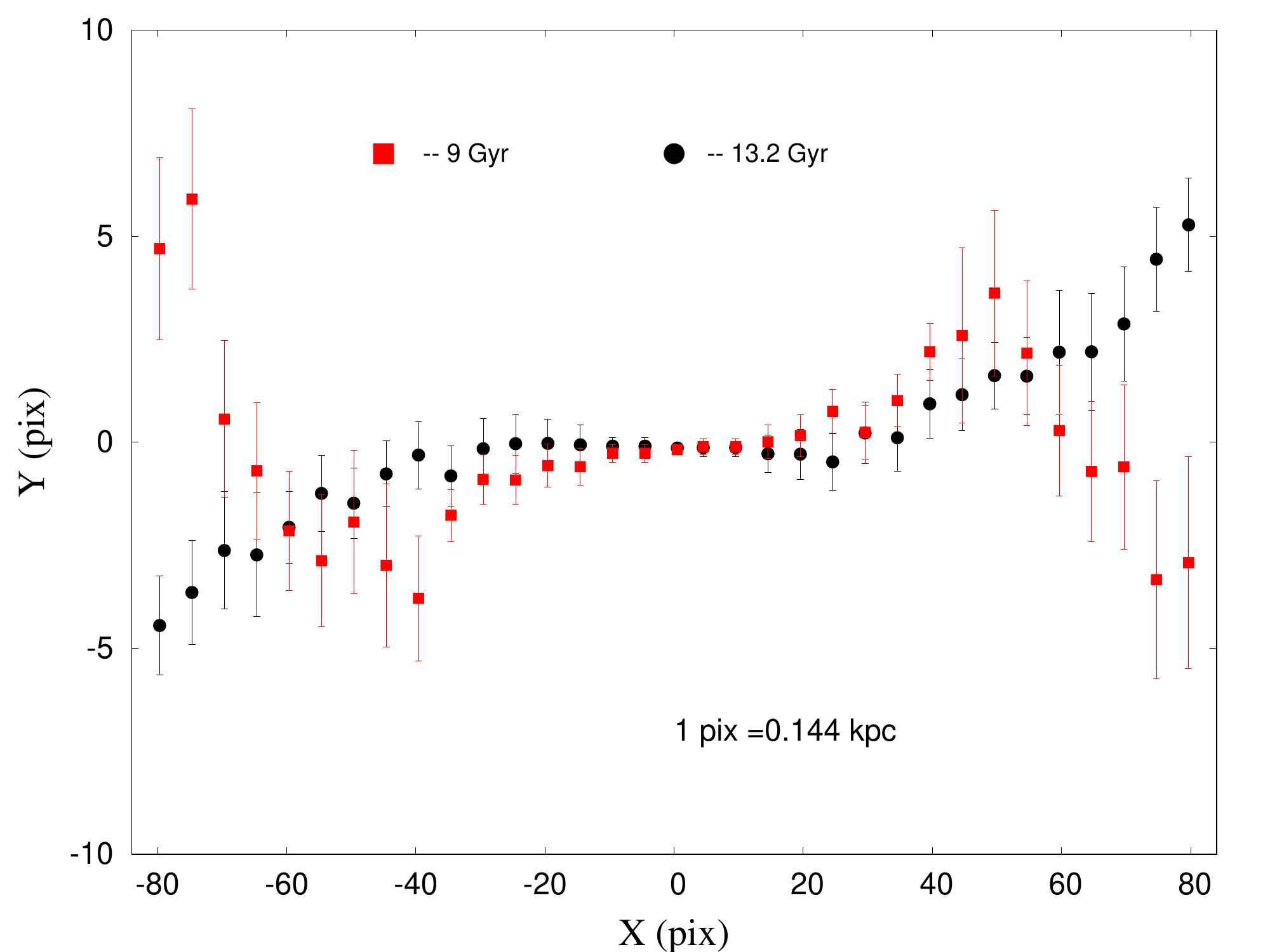}
\caption{Location of the density peaks in the $X-Y$ plane i.e., mid-plane of the disc. Error bars denote $1\sigma$ error about the mean.}
\label{fig:Barmidplane}
\end{figure}

\begin{figure*}
\includegraphics[width=\linewidth]{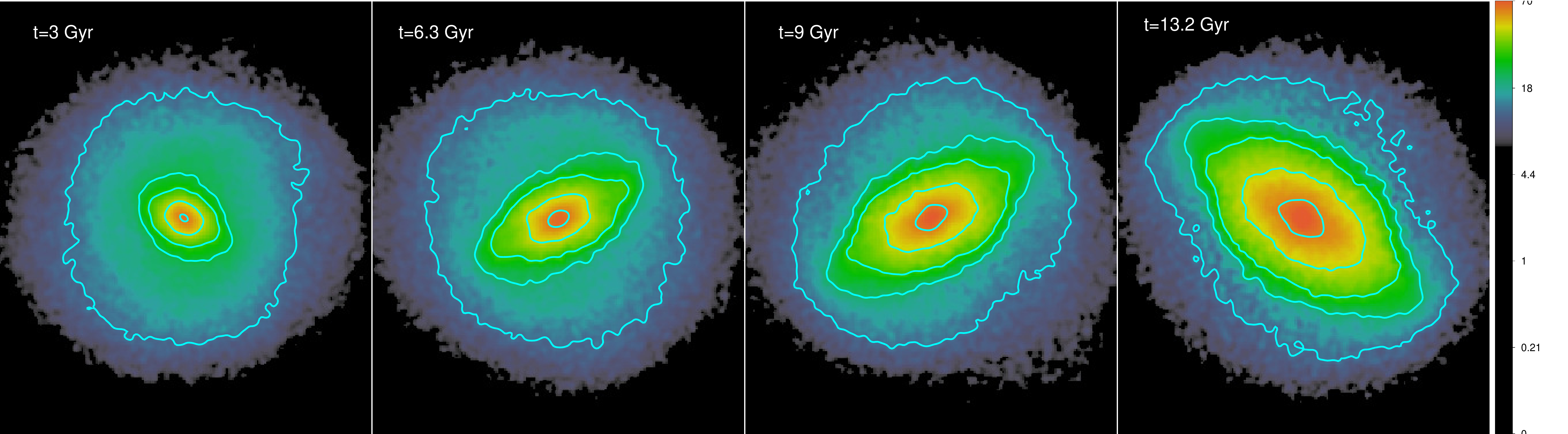}
\caption{Velocity dispersion maps are shown at four different times during the bar evolution of the 3-D peanut model. Inner to outer contours are drawn at 60, 44, 30, 23, 12 $\kms$. At $t = 9 \Gyr$, inner contour and outer bar is misaligned by $\sim 4 {\degrees}$. }
\label{fig:velocityDispersion}
\end{figure*}

In this section, we discuss the possible impact of multiple pattern speeds within the bar region. In general, one would expect to see some form of distortion in the density and velocity field if different parts of the bar rotate with different pattern speeds instead of a bar with single pattern rotation. In Fig.~\ref{fig:densitycontour}, we examine, in detail, the surface density map within the bar region at two different epochs (e.g., $9 \Gyr$ and $13.2 \Gyr$). A close inspection of the images reveals that the position angles of the density contours are not the same at all points along the bar major axis. In Fig.~\ref{fig:densitycontour}, the inner peanut region is marked by two green lines inside which the contours are well aligned. However, just outside the inner region i.e., in the intermediate region, the density contours show characteristic deviation in a systematic fashion. Further out along the bar major axis and near the ring \citep[see][]{Sahaetal2018}, which we call as the outer region of the bar or the handle of the bar, the density contours deviate in an opposite sense to the intermediate region. Based on the visual inspection, the deviation between the inner peanut region and the outer region is $\sim 3 {\degrees}$ (see Fig.~\ref{fig:densitycontour}). We quantify such deviation by modelling the density profile parallel to the minor axis. For this, we extract $20$ such density profiles on either side of the bar centre, at different locations along the bar major axis. Each of these $40$ profiles is fitted with a Gaussian function to find the location of the peak and the associated standard deviation. We utilised the IDL {\sc mpfitfun} \citep{Markwardt2009} which employs the non-linear least square method based on the Levenberg-Marquardt algorithm \citep{More1978} to carry out the fitting procedure. 

In Fig.~\ref{fig:Barmidplane}, we show the location of the density peaks on the $X-Y$ plane (mid-plane) of the galaxy. Inside the inner region (peanut), the contours are well aligned with the bar major axis (inner part). However, outside the peanut (i.e., intermediate region), the density peaks are arranged much in the well-known S-shaped warp-like fashion \citep{Sahaetal2006}. Unlike warps, this is not related to the bending of the disc midplane. In the outer region of the bar i.e., the handle of the bar, the locations of the density peaks reach a characteristic maximum and then falls back (see the case at $9 \Gyr$). Interestingly, when we compute the angle between the major axis of the inner peanut and the maximum deviation along the Y-axis (as seen in Fig.~\ref{fig:Barmidplane}), we find that the angle is around $\tan^{-1}({Y_{\rm peak}/X_{\rm peak}}) \simeq 3.4 {\degrees}$ for $t = 9 \Gyr$ and $3.5 {\degrees}$ for $t = 13.2 \Gyr$. Here, $Y_{\rm peak}$ is the maximum deviation just outside the inner peanut, and $X_{\rm peak}$ is the corresponding $X$ location. Surprisingly, this closely matches our estimate based on visual inspection of the density contours. This has also been reflected in the radial variation of the phase angle, as shown previously in Fig.~\ref{fig:phi2}.  
Similar twists of the density contours are seen in the velocity dispersion maps, as shown in Fig.~\ref{fig:velocityDispersion}. At $t = 6.3 \Gyr$ and $t = 9 \Gyr$, the outer dispersion contours are clearly offset with respect to the inner peanut. When measured, the inner and outer contours seem to be offset again by $\sim 3 - 4 {\degrees}$.
However, the same is not true at 3 Gyr when the peanut has not developed yet. One could infer that the offset between the inner and outer region occurs only when the bar enters the buckling phase. In fact, a twist in the density contours between the boxy/peanut and its outer region is observed during the buckling phase of a bar in some galaxies \citep{ErwinDebattista2016}. 
Although a slight deviation of $\sim 5 {\degrees}$ is often tolerated in defining the bar position angle \citep{Erwin2005}, we have considered them due to their systematic variation. Our current analyses show that a twist in the density and velocity dispersion contours within a long bar with peanut structure is related to differential pattern rotation of the bar. Since the offset between the inner and outer regions of the bar is small, the overall shape of the bar remains intact even when the pattern speed difference is non-negligible.

\section{Discussion}
\label{sec:discussion}
%
Previous studies of bar pattern speed assumed that the bar rotates like a rigid body with a well-defined, single pattern speed ($\Omega_{\rm p}$). The bar rotation thus introduces several well-defined resonances, e.g., Corotation, Inner and Outer Linblad resonances (ILR, OLR) in the disc of the galaxy \citep{SellwoodandWilkinson1993}. We demonstrate that the three regions in the long peanut bar rotate with different pattern speeds in the present scenario. The presence of multiple pattern speeds, as opposed to a single pattern speed, thus introduces multiple resonances (e.g., multiple CRs) in the disc. Depending on the differences in their pattern speeds, two or more CRs would either overlap or stay separated. 
Since, at later epochs (say around $t = 9 \Gyr$ and after), the differences in the pattern speed values, calculated for three regions, are not very large, the corresponding locations of CRs will not be far apart. Such a slight difference in the pattern speeds would effectively create a region containing the CRs leading to resonance overlap. This, in turn, might produce a significant fraction of chaotic orbits \citep{Contopoulos1981}.

In the past, the vital role of resonance overlap due to bar-spiral or spiral-spiral scenarios \citep[e.g., see][]{SellwoodandBinney2002,MinchevandFamaey2010,Minchevetal2011} has been investigated in the context of radial migration and reshaping the chemo-dynamical structuring of disc galaxies. The finding of this paper poses a new dynamical scenario where multiple structures encompassing the bar region rotate with different pattern speeds. It will be pretty exciting to study disc dynamics and the radial migration of stars in this scenario.

We show that different regions within the long peanut bar in our model decelerate with different rates. Past studies showed that a bar slows down eventually due to dynamical friction while transferring the angular momentum to the dark matter halo. This transfer happens preferentially at the CR \citep[e. g., see][]{Weinberg1985,HernquistWeinberg1992,DebattistaSellwood2000,Athanassoula2002,ValenzuelaandKlypin2003,SellwoodandDebattista2006,Dubinskietal2009}. Since, in our model, the pattern speed, measured in three different parts, decays with different rates, it intuitively implies that the resulting dynamical friction acting on these regions of the bar might vary as well.

Recent numerical simulations have studied the effect(s) of the slowing down of a bar in Milky Way-like systems. Using $N$-body simulations, \cite{Halleetal2018} have shown that the slowing down of a stellar bar helps stars initially trapped at the CR to be churned out in the radially outward direction.
Another study by \citet{Khoperskovetal2020} argues the radial migration of stars, initially trapped at the CR, due to the slowing down of the bar as the plausible explanation for the presence of metal-rich stars at the Solar neighbourhood. A further study by \citet{Chibaetal2021} argues that the slowing down of a bar can leave characteristic imprints (e.g., Hercules stream) in the phase-space of the stars in the Milky Way. Since different parts of the bar region decelerate at different rates in our 3-D peanut model, it would be instructive to investigate the radial migration process and the characteristic fingerprints in the phase-space of stars in such a scenario.

\section{Conclusion}
\label{sec:conclusion}
In summary, we study the evolution of a long bar in a simulated galaxy that harbours a strong peanut structure, prominent both in face-on and edge-on views. In this 3-D peanut model, we investigate how multiple structures grow with time in the bar region. We also measure the pattern speed of the long peanut bar using the TW method and characterise its temporal evolution. Our main findings are:

\begin{itemize}

\item{At later epochs ($\sim 6.1 \Gyr$ or so), the 3-D peanut model develops multiple structures encompassing the whole bar region, as identified by prominent, distinct peaks in the radial profiles of the $m=2$ Fourier coefficient. This is different from the initial single-bar scenario with a single prominent peak in the radial profiles of the $m=2$ Fourier coefficient. These structures are dynamic, changing their strength with time.}

\item{Using the TW method, we measure the pattern speeds of the bar at different regions corresponding to the locations of these multiple structures. When these multiple structures co-exist, the inner core of the long bar in our model rotates with a pattern speed that is systematically smaller compared to the intermediate and the outer regions.}

\item{The slow down of the bar follows an exponential decline. Moreover, the pattern speeds of the three regions display different decay rates. The inner region shows a faster decay than the outer regions. The decay time-scale ($\tau_{\rm bar}$) of the bar is significantly shorter for the inner peanut region ($\sim 4.5 \Gyr$) compared to the outer bar region ($\tau_{\rm bar} > 12.5 \Gyr$).}

\item{The co-existence of multiple pattern speeds in the 3-D peanut model leaves characteristic fingerprints in the density and the velocity dispersion maps of the galaxy model. The density contours in these three regions are systematically offset from one another, further reflected in the systematic modulations of density peaks about the bar major axis. Similar twists in the contours of the velocity dispersion bear the signature of multiple pattern speeds in the 3-D peanut model.}
\end{itemize}

\section*{Acknowledgement}

P.V. acknowledges IUCAA, Pune for the warm hospitality during his stay, and the DST-INSPIRE fellowship from the Government of India. S.G. acknowledges support from an Indo-French CEFIPRA project (Project No.: 5804-1).

\section*{Data availability}
The simulation data underlying this article will be shared on reasonable request to K.S. (kanak@iucaa.in).

\bibliography{main} 
\bibliographystyle{mnras}

\bsp
\label{lastpage}

\end{document}